# WEIGHT FLUCTUATIONS OF INFORMATION STORAGE MEDIA [1]

L. B. KISH[2],
*Department of Electrical and Computer Engineering, Texas A&M University, College Station, TX 77843-3128, USA*

In this essentially *Unsolved Problems of Noise* (*UPoN*) paper we further study the question recently posed in *Fluctuation and Noise Letters* (December 2007), if there is and interaction between bodies with correlated information content, and weather the observed weight transients during/after changing the information content in memory devices is due to a new type of interaction, a new type of "fifth force", or it is only a classical mechanism. We briefly discuss the issue of the great experimental uncertainty of the Newtonian gravitation constant. We also mention the peculiar experiments about sudden weight changes of humans and animals at the moment of death. The extended monitoring of four 4GB flash drives with no casing and various information content indicate a significant correlation between their weight variations and the fluctuations of ambient humidity. This is an evidence for the role of humidity and hygroscopic components, at least, for long-term weight fluctuations. A sequence of information changing experiments with such a flash drives at stable humidity conditions shows a significant variability of the transients of the absolute mass with some dependence on the information content. Finally, a related new experiment was carried out with olive oil and chilli pepper powder that was dissolved in it while the mass variations were recorded and a positive mass transient of 0.3 milligram was observed for about 10 minutes. The process represents the writing of new random information into a medium. The only classical interpretation of this mechanism would be the compression of trapped air between the grains by the surface tension of the oil, or that of in pores by capillary forces, and the resulting decrease of the Archimedes force due volume reduction.

## 1. Introduction

This paper is the continuation of our recent paper [1] about the question if there is and interaction between bodies with correlated information content. There are various challenging problems behind asking this question.

### *1.1. Anomalies of the Newtonian gravitation constant*

Most importantly, the fact that the Newtonian gravitation constant $G$ has been the least accurately known fundamental physical constant [2,3] and such an interaction, a new possibility for the hypothetical "fifth force" [4,5], may be responsible for the inaccuracy. Though Fischbach's and coworkers' fifth force hypothesis of the 1980's [5] was not found feasible, in the 1990s, several laboratories carried out new measurements with an emphasis on enhancing the accuracy of $G$ [2]. However, these efforts had backfired and the enhanced measurements met with even greater inaccuracies in the short-range value of $G$ [2], see Figure 1. In 1999, the international organization *CODATA* (Committee on Data for Science and Technology) decided [2] to increase the uncertainty of the value of the gravitational constant from 128 ppm to 1500 ppm (0.15%).

[1] http://arxiv.org/abs/0805.4175
[2] until 1999: L.B. Kiss

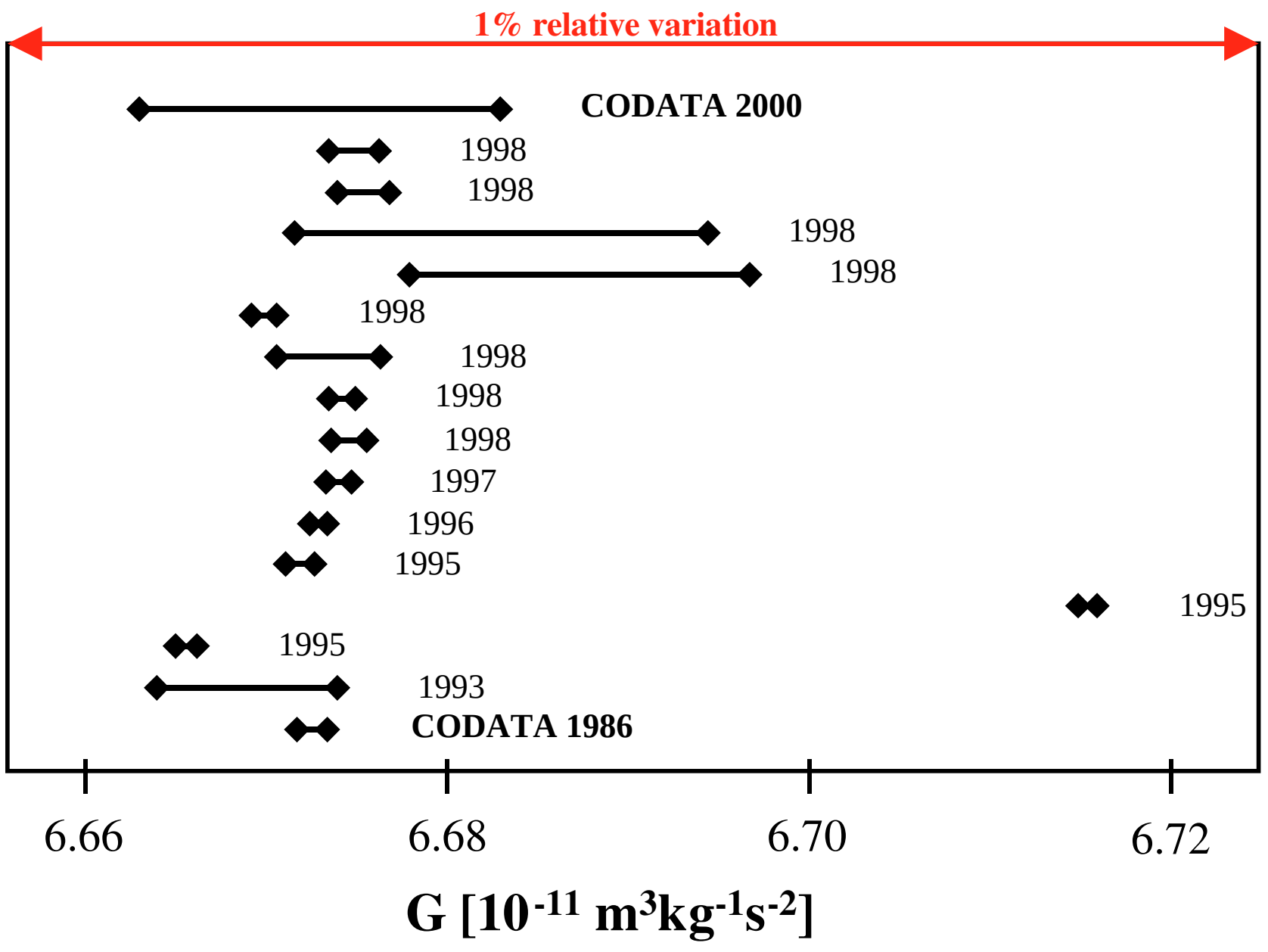

**Figure 1.** Experimental results [2] aimed to improve the accuracy Newton's gravitational constant in the period 1986-2000. The efforts resulted in the opposite implications: in year 2000, *CODATA* increased the inaccuracy range of $G$ from ~$10^{-4}$ to 1.5 $10^{-3}$ . Figure drawn by using the data in [2].

However, independently of these events, itself the original ~0.01% uncertainty is orders of magnitudes beyond that of the acceptable level of other physical constants.

### *1.2. A hypothetical new type of the fifth force*

In [1], it was hypothesized that bodies with correlated information structure may interact in a new way that is not predicted by today's physics. This new type of interaction would cause corrections of the potential energy $V_G$ of gravitation between two bodies with masses $M_1$ and $M_2$ located at distance $R$:

$$V_G = -G\frac{M_1 M_2}{R} \quad . \qquad (1)$$

The energy correction would be the function of the structural information content of the bodies, the correlation of these informations, and other parameters such as distance, pattern characteristics such as activation energy scale to maintain the pattern, size, etc. [1]. For example, let us consider the simple situation shown in Figure 2, where the bodies *A* and *B* are composed of three different components and the amount of each components is the same in *A* and *B* but the structural arrangement of the components is different. For simplicity, let us neglect all the known interactions between these bodies. Our hypothesis is that the structural information based interaction energy will be different between *A* and *A*; *B* and *B*; and *A* and *B*. If it is so, then it is also obvious that the weight of body *A* will be different from the weight of body *B* because of the different strength of the new interaction between these bodies and the structures of the earth globe.

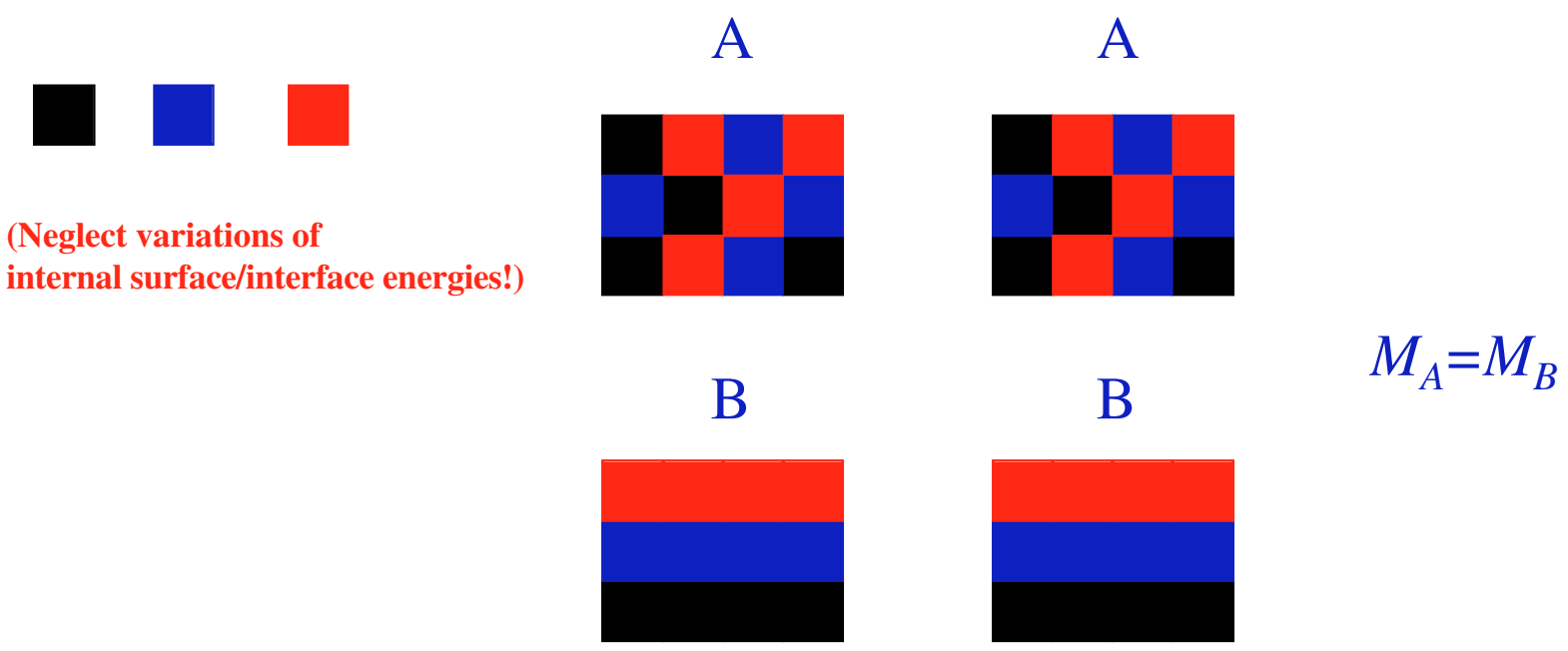

**Figure 2.** Bodies *A* and *B* have the same composition but with different structures. Does the measured gravitation-type attraction has the same strength between *A* and *A*; *B* and *B*; and *A* and *B* ? By other words, is there possibly a not yet recognized interaction related the correlation between information patterns in these bodies? If yes, that does not follow from today's laws of physics, however it could serve as a potential explanation for gravitation constant anomalies.

Without any experimental knowledge of this situation it is difficult to construct a theory because of the too many possibilities. However the implications of such a hypothesis are very serious. Without the aim of doing a complete analysis, we illustrated in [1] that the *energy conservation law* would have an impact on the energy need of creating such a structure and the strength and range of interaction in practical structures.

As an ad-hoc example [1], we used the *mutual information* $I_{1,2}$ of two digital information structures to illustrate how such an interaction may look like

$$I_{1,2} = \sum_{i,j} \sum_{x_i=H,L} \sum_{y_j=H,L} p_{1,2}(x_i,y_j) \log\left[\frac{p_{1,2}(x_i,y_j)}{p_1(x_i)p_2(y_j)}\right] \tag{1}$$

where $p_1(x_i)$ is the probability density function that the i-th element $x_i$ of *pattern-1* belonging to *body-1* has a given value (high or low, $x=H$ or $x=L$, respectively); similarly $p_2(y_j)$ is the probability density function that the j-th element $y_j$ of *pattern-2* belonging to *body-2* is a given value ($y=H$ or $y=L$, respectively); and $p_{1,2}(x,y)$ is the joint probability function of these situations. We pointed out [1] that the square of $I_{1,2}$ has properties resembling to gravitation because by multiplying the size and mass $M_1$ and $M_2$ of the bodies while repeating the same specific patterns in them will cause $I_{1,2}^2$ to grow proportionally to the product $M_1M_2$ , that is, $I_{1,2}^2$ scales proportionally with the gravitational force between the two bodies. As a possible example, we introduced an interaction potential, which is a correcting energy term to the gravitation energy given by Equation 1, in the following form

$$V_{inf} = \frac{\alpha}{R} I_{1,2}^2 \quad , \tag{3}$$

where the $\alpha$ is the function of the combination of physical parameters of the structure (see examples above).

### *1.3. Measurements of the static interaction*

We carried out the experimental tests [1] with a HR-202i precision scale, see Figure 3. No interaction between flash memories with the same information was detected at the given weight resolution (10 microgram). In the experiments one flash drive filled up with white noise was placed on the scale and its weight was determined. Then another flash drive with the same information content was placed over it at a distance of 1 cm. Possible changes in the weight of the first flash drive were then monitored. We typically observed a small weight transient (in the order of 0.03) mg, which gradually relaxed toward zero within a time period of a minute, and we interpreted this effect as an artifact caused by the thermal disturbance of the scale's internal atmosphere when we opened its door and put the other flash drive there. Using the negative result and our resolution in Equation 3 we obtained [1]:

$$\alpha < 5.21 * 10^{-33} \frac{N}{bit^2 m^2} \quad . \tag{4}$$

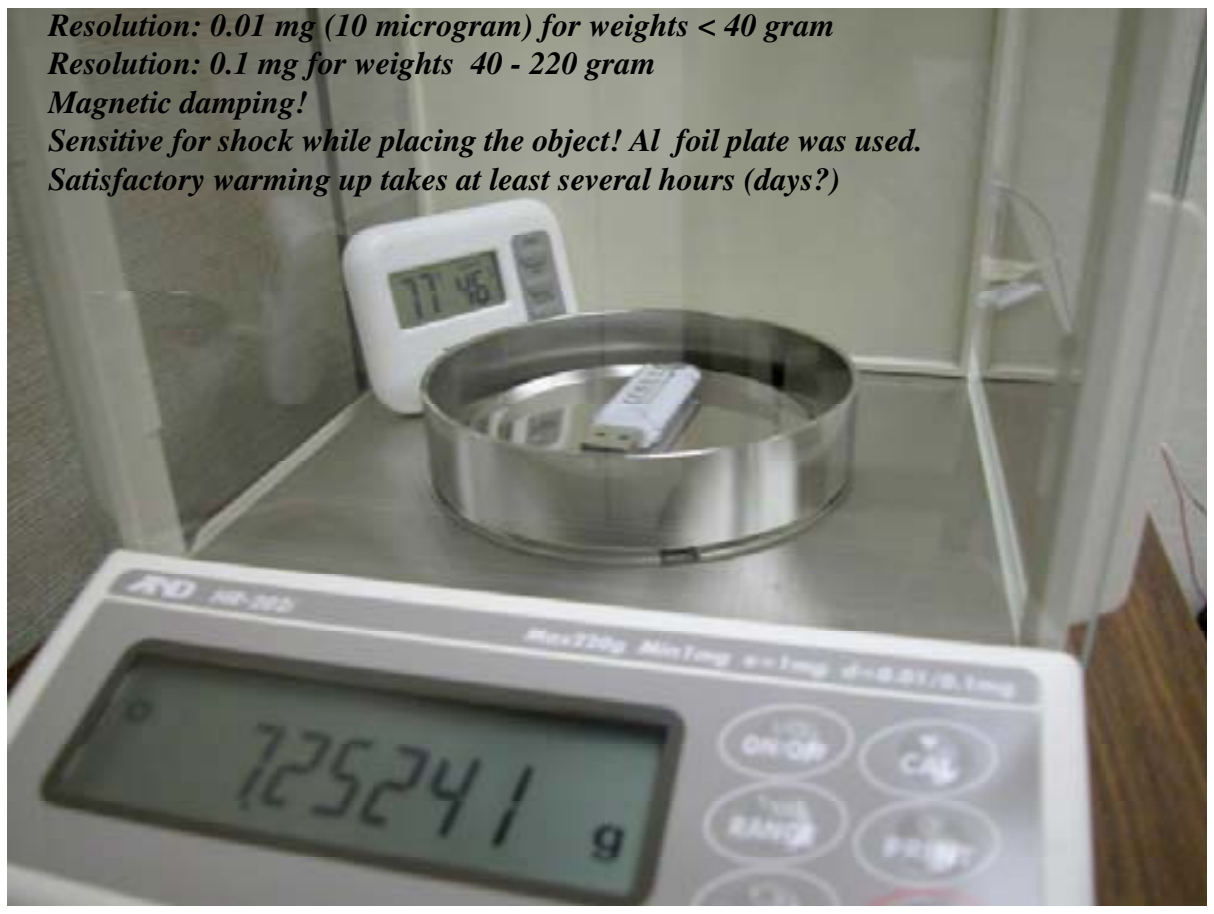

**Figure 3.** The scale, the humidity and temperature gauge, and a 2 GB flash drive. For more details see the text in the figure. At the experiments reported here the flash was placed on the scale together with a handmade *Al* foil plate which helped to reduce the shocking of the scale thus the induced offset was negligible.

Finally, note: let us suppose the interaction exists and it does explain part of the gravitational anomalies. Is it then surprising that we have not seen any interaction? The answer is *no*. We have some $10^{11}$ bits in a flash drive and some $10^{23}$ atoms. Even if we scramble these $10^{11}$ bits, the relative structural change in the system is extremely low compared the structural pattern differences even between identically looking samples of pure single crystals. Our system cannot even detect the gravitation interaction between two flash drives, thus it cannot see the correction of gravitation forces either. Still, we had a vague hope that two absolutely correlated $10^{10}$ bit random structures may show some visible interaction due to a "resonance" effect, especially that this kind of absolute correlation between random structures is non-existent in nature, thus the experiment addressed something new that was never tested.

### *1.4. Observation of unexplained transient weight changes in memory devices*

Originally we expected only at most the static interaction outline above however, due to the observed spurious experimental results, unexplained weight variations, see Figure 4 as an example from paper [1], we expanded the picture with the dynamical interaction idea:

$$V_{trans} = V\left[\frac{dI_{1,2}}{dt}, \alpha, R, t\right] \quad , \tag{5}$$

where the transient is caused by the changing of correlation between the information patterns and depending on the conditions, it can show various relaxation times constants.

Figure 4 shows the time relaxations of weight loss transients experienced in DVD disks [1]. The relaxation is faster with lower humidity which seems to contradict to simple explanations based on loss of water from hygroscopic components due to the elevated temperature during writing [1].

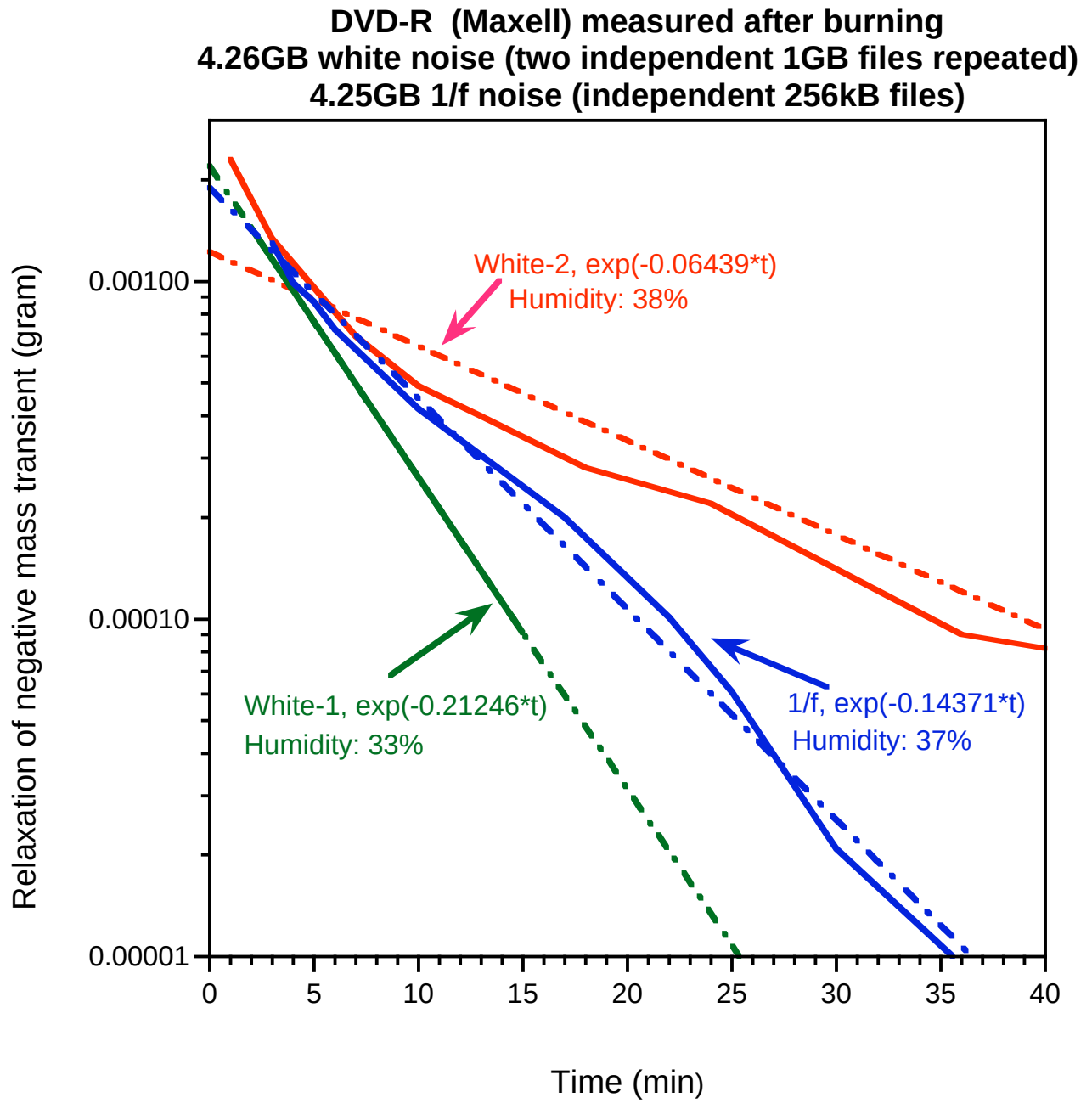

**Figure 4.** Semi logarithmic plot of the absolute value of the negative mass transients in DVD-R media [1]. A straight line represents an exponential decay. Solid line: measurements; dashed line: exponential fit. The shortest relaxation time constant is about 5 minute and even that is much longer than the estimated thermal relaxation time (0.5-1 minute). The longest relaxation time constant is about 15 minutes.

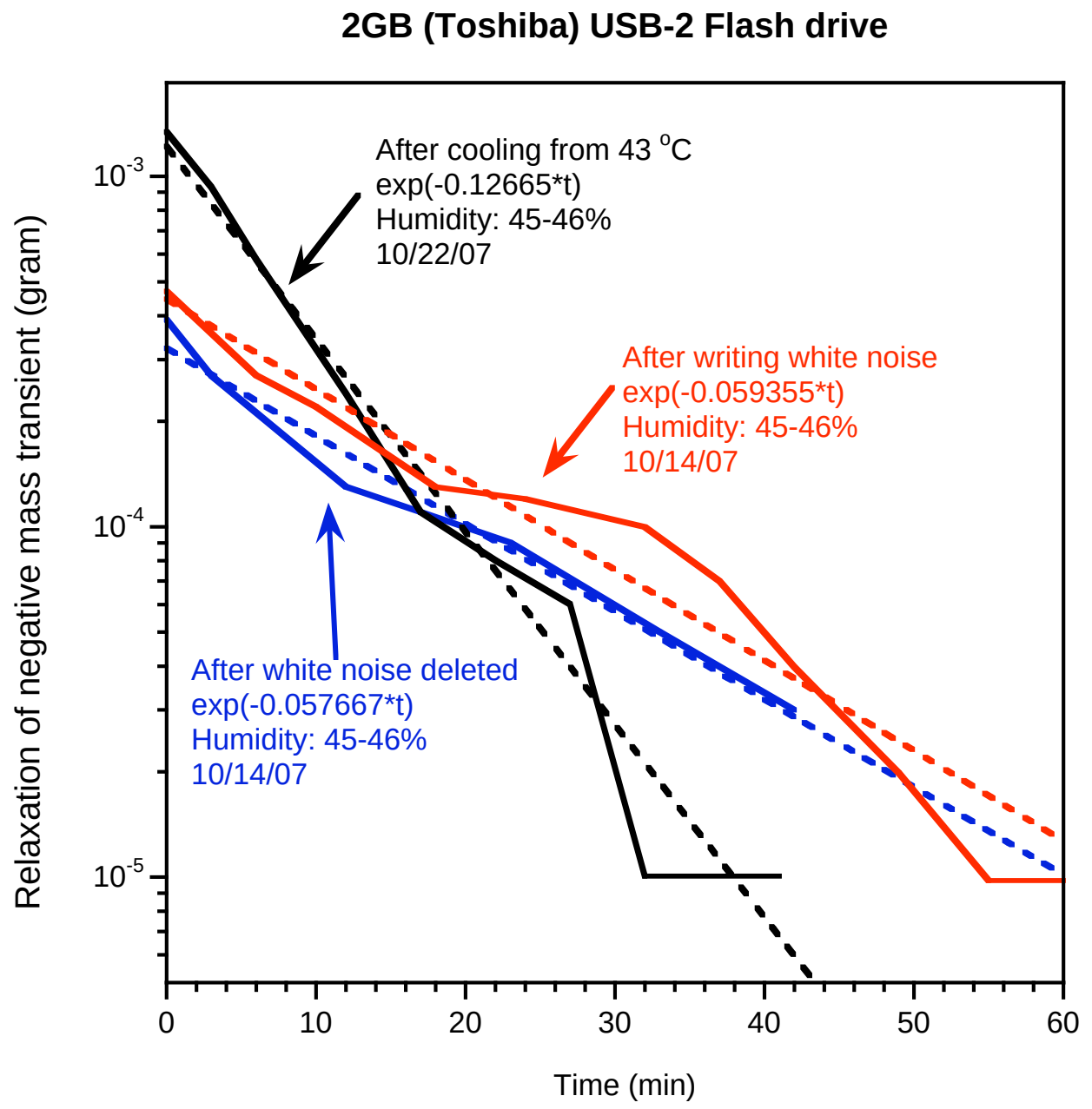

**Figure 5.** Semi logarithmic plot of the absolute value of the negative mass transients measured after the thermal bath (see Fig. 7) and after information change at different conditions are shown for comparison. A straight line represents an exponential decay. Solid line: measurements; dashed line: exponential fit. The relaxation time constant after the thermal bath is about half of the relaxation time observed after recording or deleting white noise from the device.

These observed relaxations of the negative mass transient in a 2GB flash drive [1] have significantly different dynamics compared to transients observed during cooling after stationary external heating of a flash drive [1], see Figure 5. An *MP3* player has also suffered comparable mass loss during playing music, even though its information content was not changed. We discussed in [1] that simple classical interpretation of the observed weight transients could potentially be absorbed water in hygroscopic components or lifting Bernoulli force due to convection of warm air, however comparison of relaxation time constants at the various measurements did not seem support an obvious explanation in this fashion. If further studies with different scales and/or better controlling of the experimental conditions provide a quantitative classical explanation for the transient weight effects, the transient picture should be removed from the hypothesis. However, if the information-dependence of the transient force is confirmed, then it may mean that writing, reading or deleting the information in any of these devices causes peculiar weight transients, which is followed by various types of relaxation processes [1]. Note, the weight transients are sometimes positive; see the olive oil experiment in Section 2.3.

### *1.5. Unexplained weight transients of humans and animals at death*

We mention this issue with some hesitation due to the concerns about potential animal experiments and due to the complexity of biological systems, however we think it is appropriate to mention these findings about sudden unexplained weight changes of humans and animals observed at death ([6,7] and references in [7]). In humans, in a small sample of volunteers, a sudden negative mass change of 10-21 gram was documented at death [6], while in various types of animals, a positive weight transient of 18-780 gram with duration of 1-6 second (with scale time constant of 0.2 second) was observed [7]. An objective critical analysis of the situation of the experiments in [6] is given by Mary Roach [8,9] and she also arrives at the conclusion that these experiments look serious. The main reason why we mention these works is that we have observed similar types of (negative and positive) weight changes when we changed the information in storage media, see above and in Section 2, and because these papers seem to be careful works at the particular conditions with the available resources.

In the rest of this paper we report the new experimental results obtained after publishing paper [1].

## 2. The new experiments

### *2.1. Long-time correlations between flash drive weight and air humidity*

CDs are known to be hygroscopic but DVDs are considered to be less so. They but have much larger surfaces than a flash drive. After removing the casing of flash drives, see Figure 6, it is obvious that the surfaces which may warm up during writing and relatively small, about 100 times less than the surface of CDs and DVDs. Thus, the effect of absorbed water by hygroscopic components may not seem to be a reasonable explanation of the observed negative weight transients at information exchange. However, our new experiments with 4GB Verbatim flash drive seems to contradict to this assumption.

Over a period of 54 days, we had recorded the weight fluctuations of four flash drives of this kind (Figure 6) with various types of information patterns. Figure 7 shows the recorded data which clearly show correlations between the air humidity and longterm weight fluctuations.

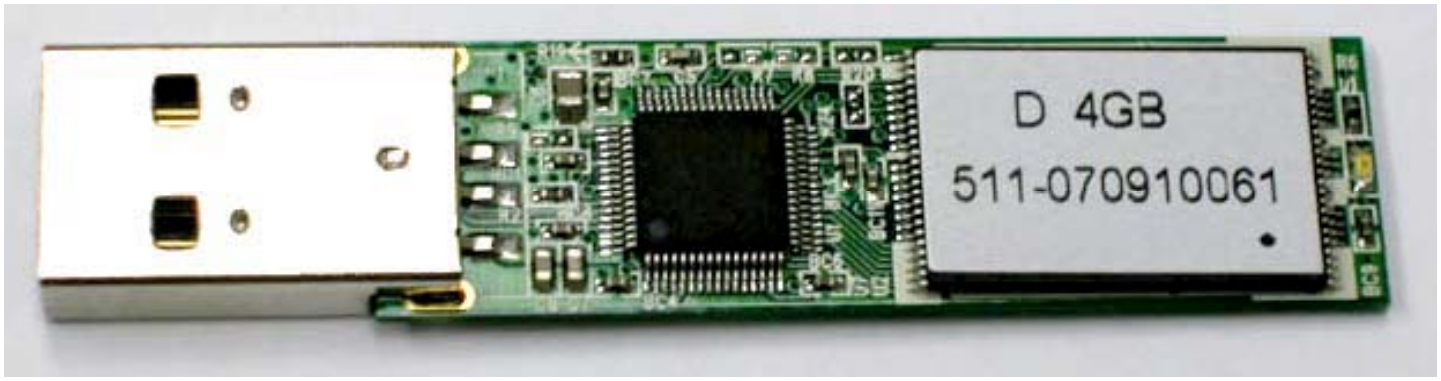

**Figure 6.** Verbatim 4GB flash drive with casing removed for the experiments reported in this paper.

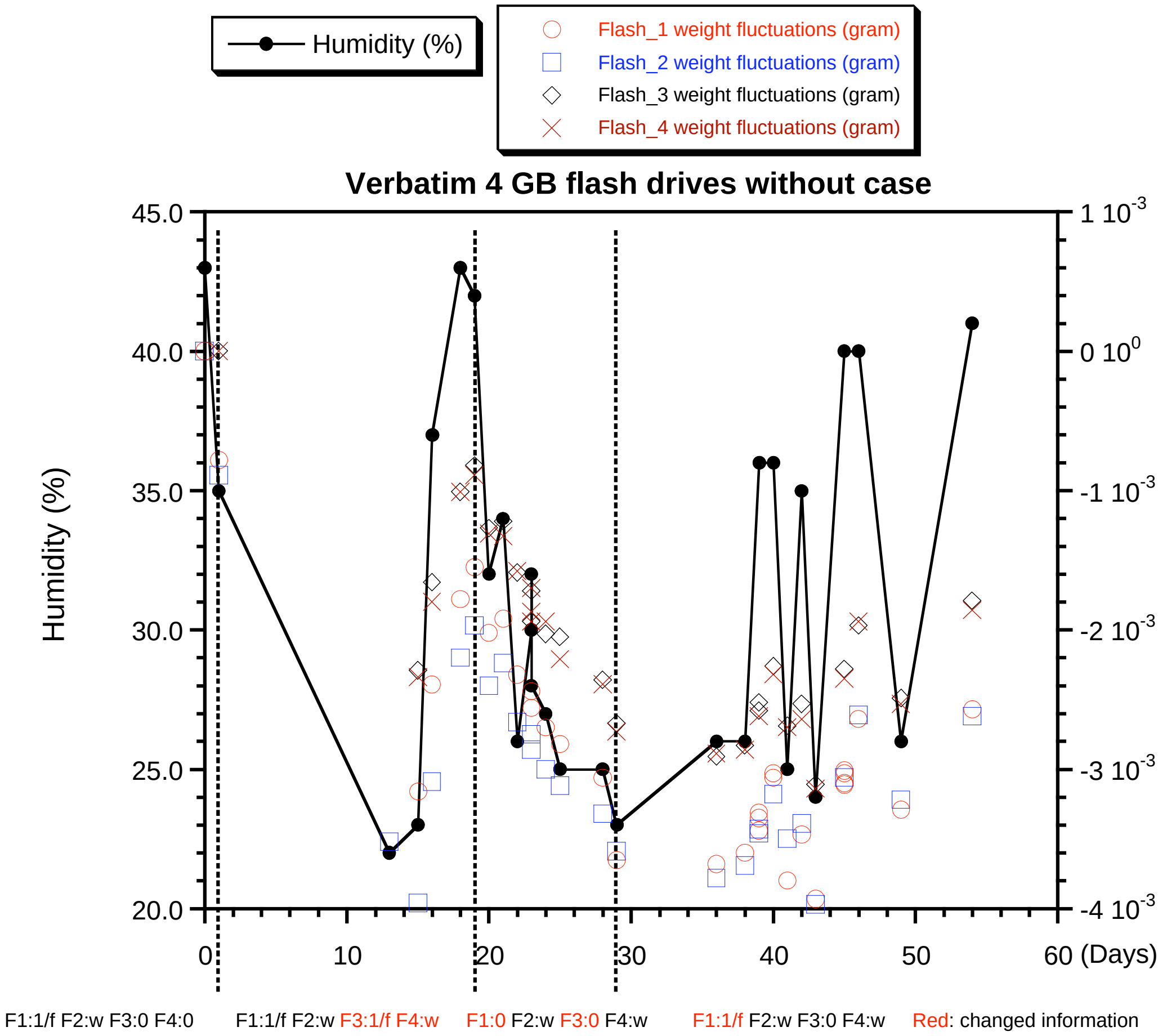

**Figure 7.** Correlations between the air humidity and longterm weight fluctuations in four 4GB Verbatim flash drives without casing (see Figure 6). The bottom line shows the type of information pattern used within the time period between the vertical dashed lines. The change of information was done at the beginning of the period)

In Figure 8, the time derivative of these data is given. Because of the unknown details between measurement points, the real information here is between the signs of the time derivative of the humidity and weight fluctuations (data below or above the dashed line), which shows strong correlations. The weight data relevant for different pattern types in Figures 7 and 8 do not show pattern-characteristic dependence. In conclusion, Figures 7 and 8 serve with evidence that at least the long-term component of weight fluctuations can be related to humidity captured in the hygroscopic components of the drives. Weight transient measurements in dry air atmosphere may serve crucial data to evaluate the role of humidity. In dry air, no relaxation should be observed if the transients observed earlier are due to the loss and regaining of water.

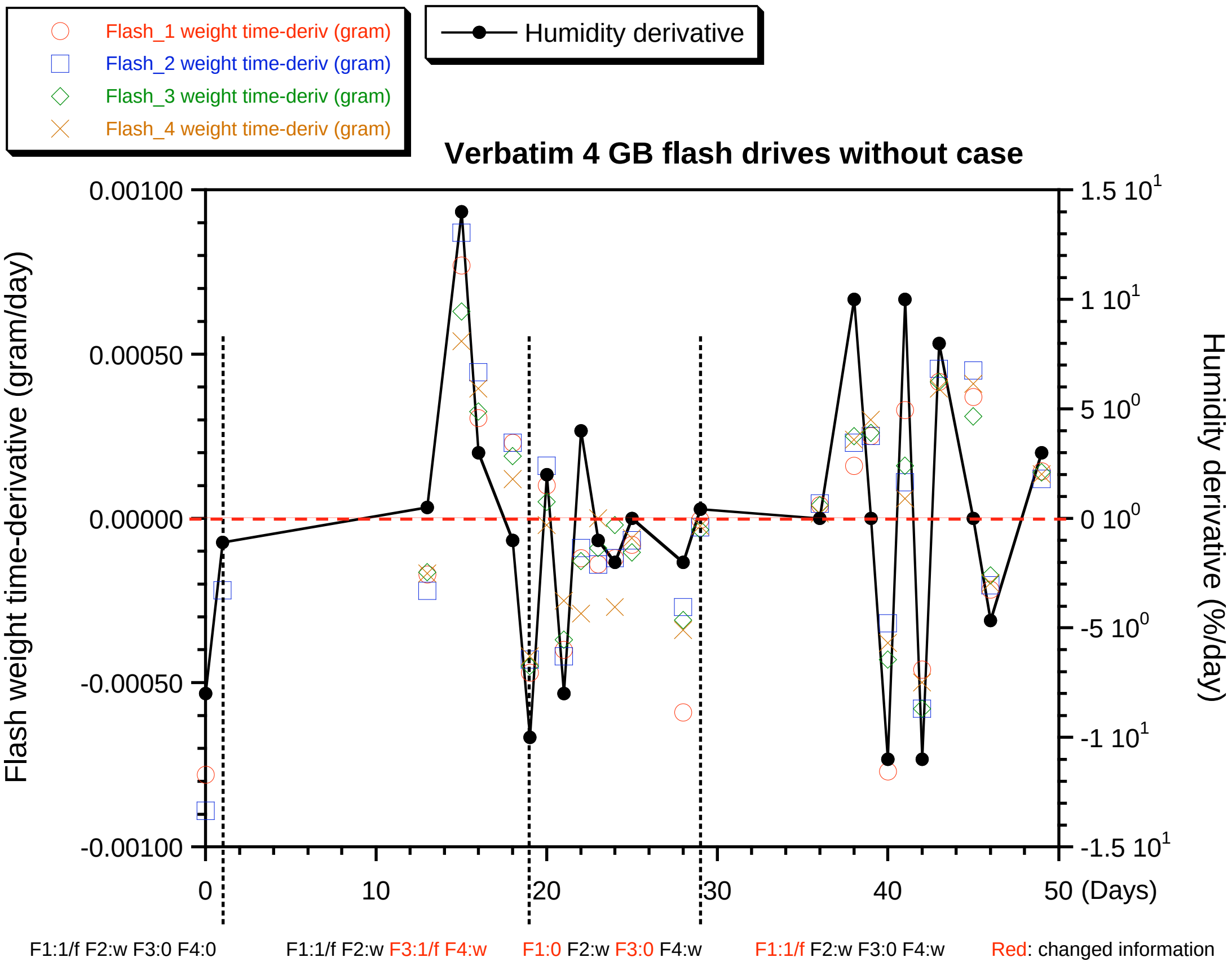

**Figure 8.** Correlations between the sign of the time-derivative of the air humidity and that of the weight fluctuations.

## *2.2. Sequential experimental series with a flash drive at the same ambient conditions*

To reduce uncertainties due to the long-term weight fluctuation components induced by slow humidity variations, we run a quick sequential experimental series with one of the 4GB Verbatim flash drives with no casing. The information was changed, the weight relaxation measured, and then new information change and weight measurements followed resulting eight measurement sequences. The whole process took less than 4 hours while the temperature and humidity in the room were stable. The results are shown in Figures 9 and 10. In Figure 9, the short-range relaxation of negative weight transient is shown. The relaxation was considered as "finished" when the weight change in 3 minutes was less than the resolution (10 microgram). Due to the long-term components mentioned above, this may introduce an arbitrariness in this kind of evaluation thus the time evolution of the absolute mass of the drive may be more accurate tool to monitor the different behaviors, see Figure 10.

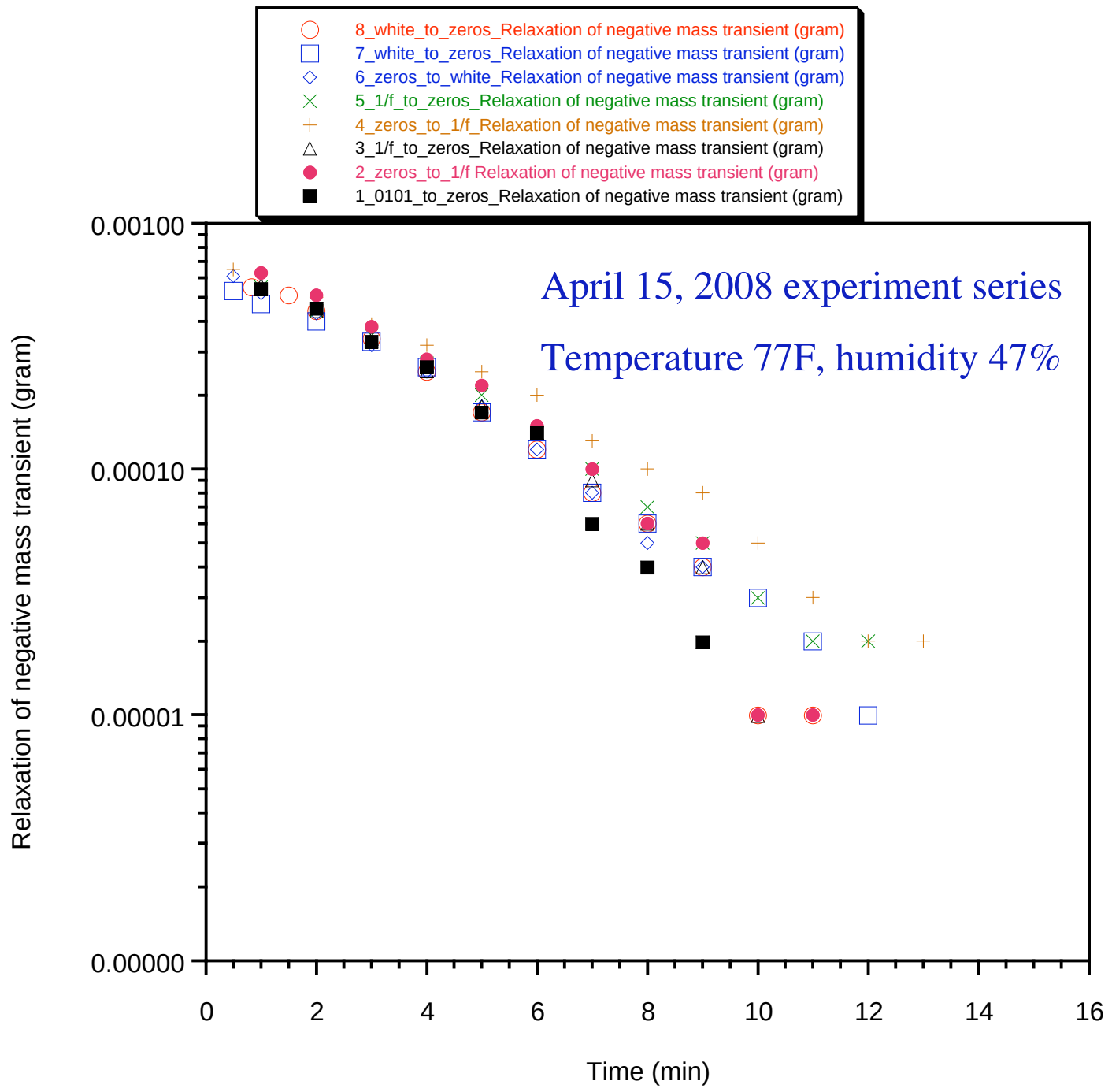

**Figure 9.** Sequential experimental series to study the relaxation transient after information change. Because the zero point has some arbitrariness, on the next figure the absolute mass transients during the same measurements are shown. The numbering is in the chronological order of the measurements.

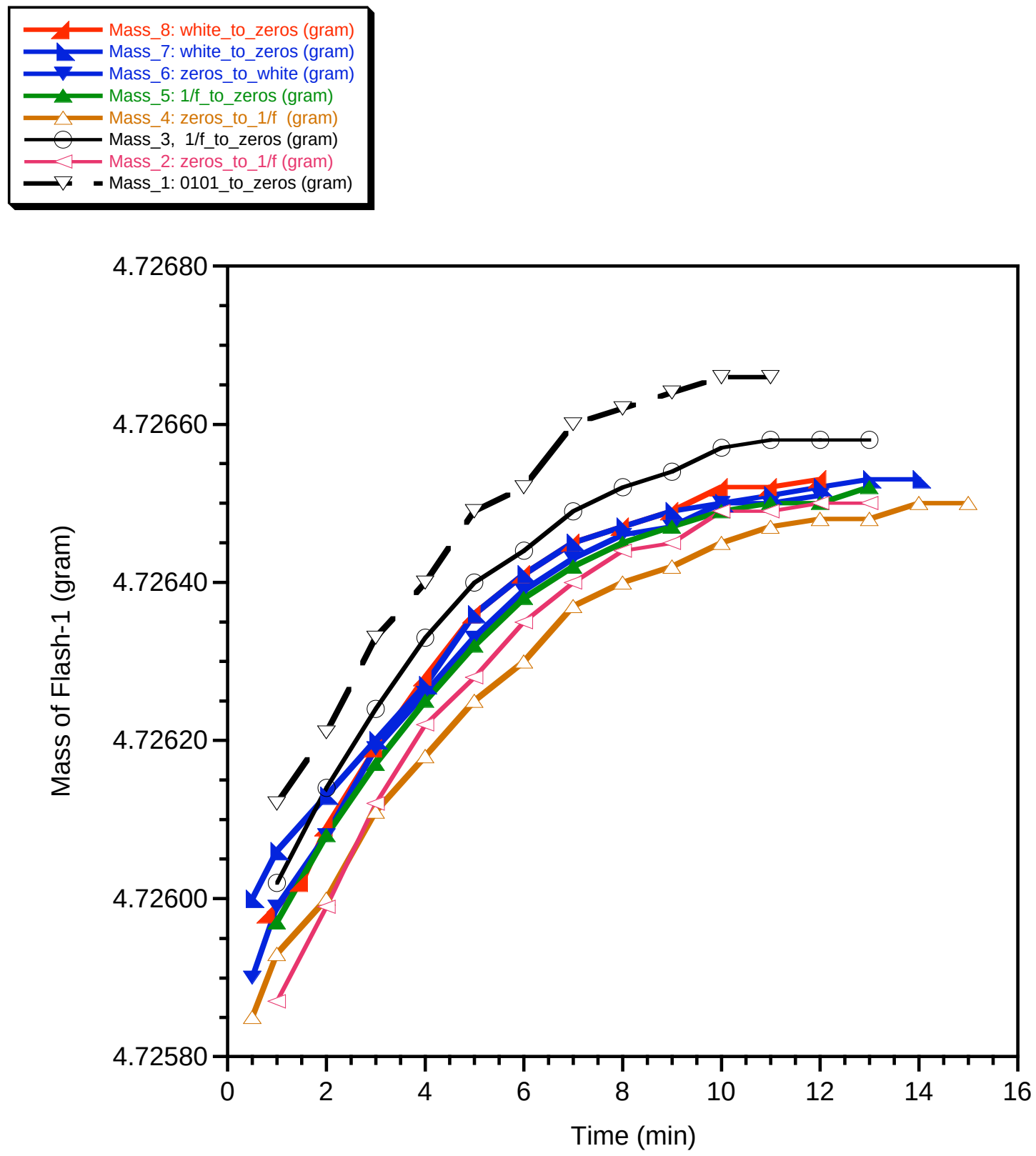

**Figure 10.** Sequential experimental series with the absolute mass transients. The numbering is in the chronological order of the measurements. The whole measurement series took less than four hours.

It is obvious from Figure 10 that there are large variations in the behavior. In the dynamical interaction fashion the reason would possible be the varying interacting structures within the earth globe and the environment. However, one can also give some partial explanation classically by using the long-term effects due to losing water and using the facts that the writing of different types of information patterns into the flash drive takes different times due to the way of data handling by flash drives. The ratios of time durations needed to write 1/f noise, white noise and zeros were about 5:2:1. Thus the fact that the 1/f noise related curves are at the lowest position in Figure 10 may be plausible however the finer details and different relaxation speeds are not explained by this simple observation.

### *2.3. Positive mass transient at "writing information" into a media in a natural way*

Finally we tested a new way of writing information into a medium. Dissolving a solid body in a fluid virtually places "white noise" pattern in the fluid medium. We tried such an experiment earlier with water and sugar however weight losses due to the fast evaporation of water were interfering with the observed effects and we had to abandon the experiments.

This time we used olive oil (extra virgin, cold pressed) as fluid medium which has a low vapor pressure and slow evaporation rate. The solid particles for "writing" the information into the oil medium were chilli pepper powder mix (Sam's club). We used the well know fact that a significant portion of chilli pepper gets dissolved in cooking oil. We placed 40 $cm^3$ olive oil on the scale and waited until the weight stabilized. Then we placed and 4.3 gram chilli powder above it in an Al foil plate, see Figure 11 (left). We monitored the weight of the system which was linearly decreasing due to evaporating the water from the chilli powder which was quite agglomerated due to its water content. Then we poured the chilli powder into the oil and further monitored the mass. The total mass executed a well-defined positive mass transient of 0.3 milligram with duration in the order of 10 minutes, see Figure 12.

One possible interpretation could be the dynamical interaction of varying information pattern with patterns in the earth globe and in the environment, similarly to how this interpretation could be used to explain the observed time dependence of gravitation anomalies at some of those experiments [2].

Concerning a classical explanation, the only possibility we seem to have is to suppose a transiently reduced total volume during the dissolution and the corresponding effect of transiently reduced Archimedes force. However, the needed volume reduction to explain this mass change is huge, 0.3 $cm^3$, and that is not easy to explain. One effect at hand seems to be to suppose that air bubbles captured between chilli grains are compressed by the surface tension of olive oil. Using the surface tension value of olive oil (0.032 N/m) we can calculate that an air bubble with 10 micron radius will have 0.06 atmosphere excess pressure, that is about -6% relative volume change; and at this situation it needs about 5 $cm^3$ captured air to explain the effect. Similar possibility is air compression by the oil via capillary forces in pores. Further experiments with other materials, especially where captured air bubbles and air in pores can be excluded from the explanation are needed to clarify the source of this effect.

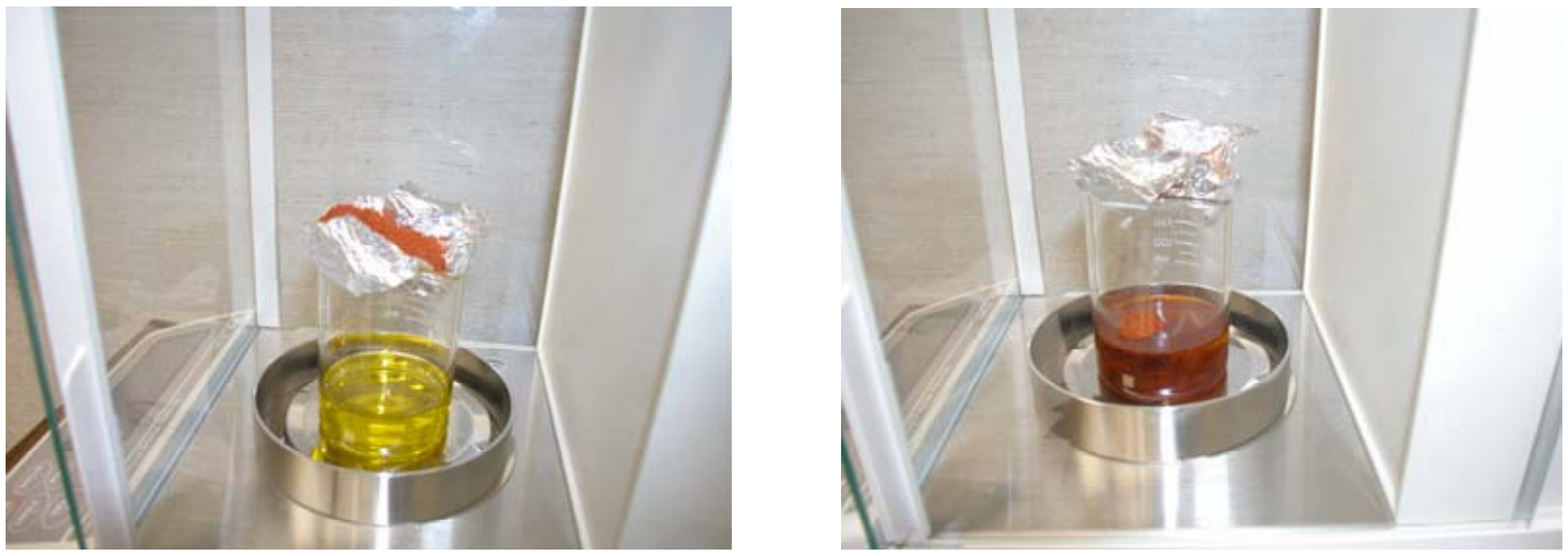

**Figure 11**. "Writing information" into a media in a natural way: dissolving chilli pepper in olive oil. Left figure: initial total mass measurement of the 40 cm$^3$ extra virgin (cold press) olive oil in the glass, and 4.3 gram chilli powder above it in the Al foil plate. Right figure: total mass measurements during "writing". See Figure 12.

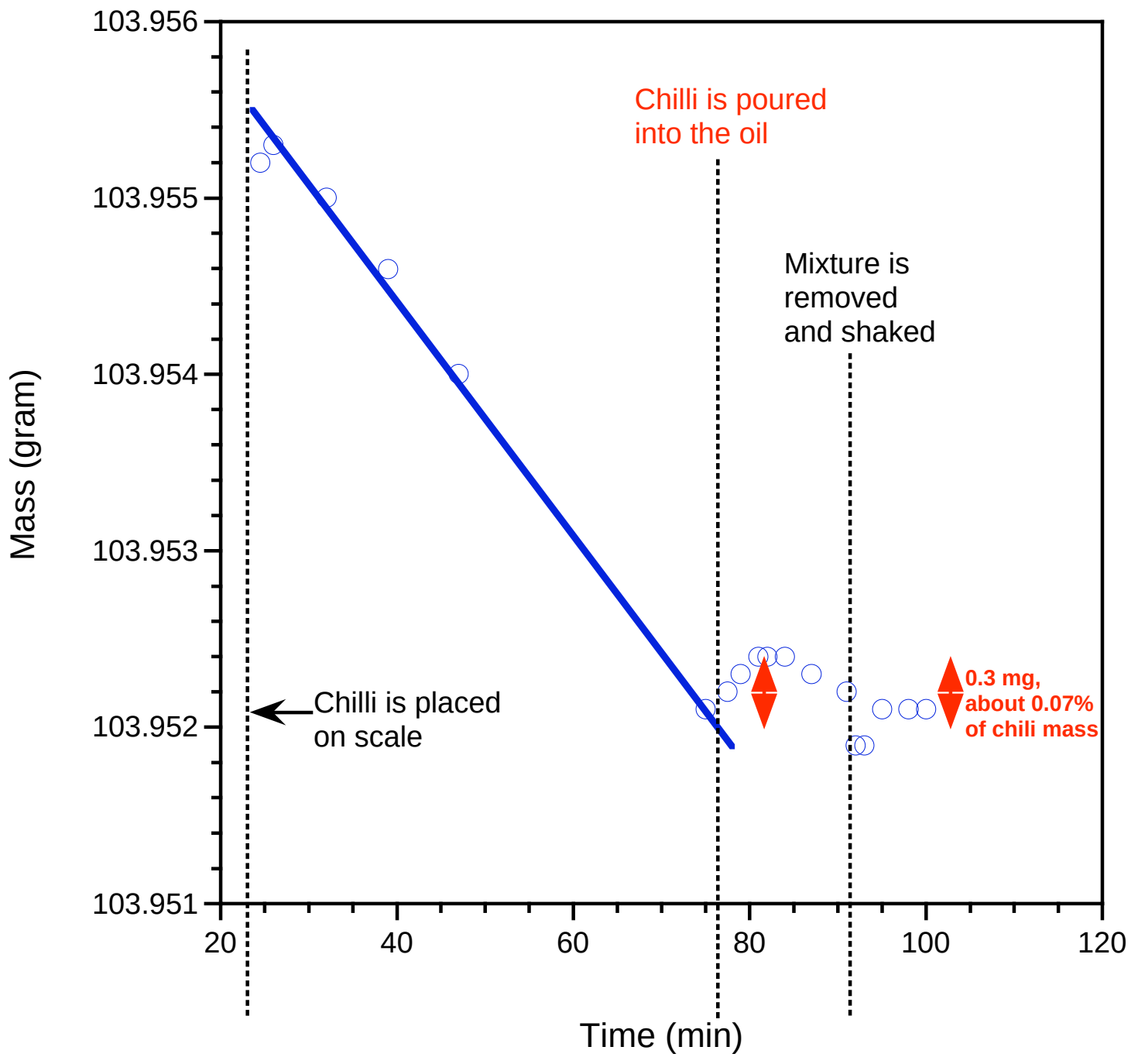

**Figure 12.** The total mass of the system shown on Figure 11. Before the chilli powder is poured into the oil, it gradually lost weight due to the evaporation of stored water. However during dissolving it, a positive weight transient was observed. During repeating the experiment, the same effect was observed.

## 3. Conclusions

The described phenomena, observations and ideas in this paper are real "Unsolved Problems of Noise". More controlled experiments with new ideas and/or better resources may clarify the observed effects. However, concerning the possible interaction between information structures, even if experiments with certain sensitivity show no interaction effect, the interactions between information patterns may exist at a lower strength below the sensitivity limit of the actual measurements.

## Acknowledgements

A discussion with Krishna Narayanan, Arun Srinivasa, Zoltan Gingl, Peter Heszler, Derek Abbott and Peter Makra is appreciated.